\documentclass[sigchi]{acmart}
\renewcommand\footnotetextcopyrightpermission[1]{}
\AtBeginDocument{%
  \providecommand\BibTeX{{%
    \normalfont B\kern-0.5em{\scshape i\kern-0.25em b}\kern-0.8em\TeX}}}

\setcopyright{acmcopyright}
\copyrightyear{2023}
\acmYear{2023}
\acmDOI{XXXXXXX.XXXXXXX}

\acmConference[Conference acronym 'XX]{Make sure to enter the correct
  conference title from your rights confirmation emai}{June 03--05,
  2018}{Woodstock, NY}
\begin{document}

%%\title{Multimodal Conditional Urban Street Layout Design}
\title{Street Layout Design via Conditional Adversarial Learning}
% \author{Lehao Yang}
% \email{george.yanglehao@zju.edu.cn}
% \orcid{0009-0009-1559-2102}

% \affiliation{%
%   \institution{Zhejiang University}
%   \streetaddress{38 Zheda Rd}
%   \city{Hangzhou}
%   \state{Zhejiang}
%   \country{China}
%   \postcode{310027}
% }

\author{Lehao Yang, Long Li, Qihao Chen, Jiling Zhang, Tian Feng, Wei Zhang}
\affiliation{%
  \institution{Zhejiang University}
  \streetaddress{38 Zheda Rd}
  \city{Hangzhou}
  \state{Zhejiang}
  \country{China}
  \postcode{310027}
}

\renewcommand{\shortauthors}{Yang et al.}

\begin{abstract}
Designing high-quality urban street layouts has long been in high demand, but entangles notable challenges. Conventional methods based on deep generative models are yet to fill the gap on integrating both natural and socioeconomic factors in the design loop. In this paper, we propose a novel urban street layout design method based on conditional adversarial learning. Specifically, a conditional generative adversarial network trained on a real-world dataset synthesizes street layout images from the feature map, into which an autoencoder fuses a set of natural and socioeconomic data for a region of interest; The following extraction module generates high-quality street layout graphs corresponding to the synthesized images. Experiments and evaluations suggest that the proposed method outputs various urban street layouts that are visually and structurally alike their real-world counterparts, which can be used to support the creation of high-quality urban virtual environments.
\end{abstract}





\begin{teaserfigure}
  \includegraphics[width=\textwidth]{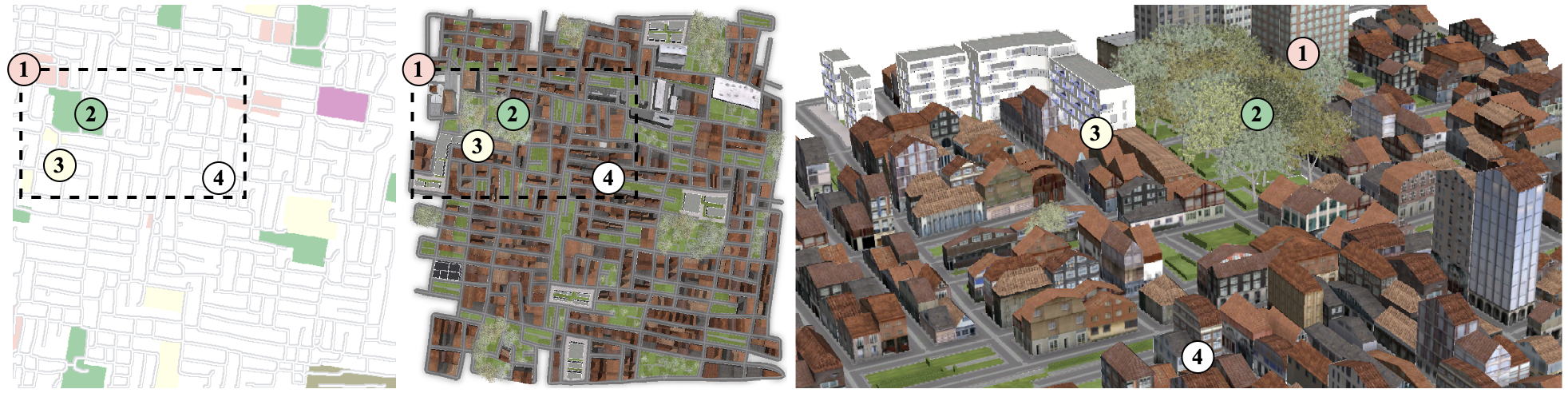}
  \caption{An urban street layout designed by our method (left), and its corresponding three-dimensional virtual environment (middle: top view; right: isometric view). Land use types are identified (1: commercial; 2: parkland; 3: education; 4: residential).}
  % \Description{Enjoying the baseball game from the third-base
  % seats. Ichiro Suzuki preparing to bat.}
  \label{fig:teaser}
\end{teaserfigure}

\received{20 February 2007}
\received[revised]{12 March 2009}
\received[accepted]{5 June 2009}

\maketitle

\section{Introduction}
Cities are where massive people live and socialize in the real world. An ineffective urban street network limits a city' capability to support smooth travel and transport, enabling street layout design to become a critical task in the area of urban planning. Meanwhile, the gaming industry has also regarded street layout design as an emerging topic in recent years, for visually and structurally sound street networks contribute to immersive and engaging user experience in the virtual urban environments. Being essential to the above-mentioned fields, this challenging task requires professionals to contemplate various factors towards high-quality street layouts given the complexity of cities~\cite{1984_Sheffi}.

% Cities are where the majority of people live and socialize in the real world. Without an effective urban street network, a city can hardly enable its residents to travel or function smoothly. Urban street layout design is a critical aspect of urban planning and design. Street layout design must account for a variety of factors, including land use, traffic patterns, safety, and accessibility, with the ultimate goal of creating a functional and efficient road network that serves the needs of pedestrians, cyclists, and motorists alike. Beyond the realm of urban planning, the gaming industry has also recognized the importance of street layout design in creating immersive and engaging game worlds. A well-designed street network can make it easier for players to navigate the city and complete objectives, while a poorly designed one can lead to frustration and confusion. With the instantaneous generation of game scenes, efficient and diverse street layout designs are necessary to support this game mechanic and provide a richer gaming experience. In conclusion, street layout design is an essential component of both urban planning and the gaming industry, requiring careful consideration of a wide range of factors to ensure the creation of a functional and engaging cityscape.

The process of street layout design is often time-consuming and tedious, requiring significant knowledge and expertise. Intensive attempts have somewhat allowed users to focus on high-level goals in street layout design. For example, the rule-based methods~\cite{2001_Parish,2012b_Vanegas,2014_Benes} adopt procedural models for quick generation; the example-based methods~\cite{2008_Aliaga,2012_Yu,2016_Nishida} rely on structurally decomposing and statistically analysing real-world data; and the optimisation-based~\cite{2013_Yang,2016_Peng,2016_Feng} encapsulate the task in an optimization process being subject to different costs. These traditional methods accomplish the task driven by natural and socioeconomic factors closely related to city development, but are yet to significantly disengage user intervention (e.g., rules, parameters, and goals) from the process.

The revolutionary development of deep learning techniques has enabled automatic generation of customized high-quality street layouts under less user intervention. Specifically, deep generative models extract structural and textural features from a large amount of data on real-world street networks, and fuse these features for street layout design. For example, generative adversarial network (GAN)~\cite{2020_Goodfellow} and its variants~\cite{2014_Mirza,2017_Isola} can be trained to map a random noise to street layouts replicating the style of the original patch~\cite{2017_Hartmann}, or synthesize street layouts being aware of user-given conditions (e.g., sketches, junction guidance, and topographic information)~\cite{2020_Fang,2021_Fang,2022_Fang}; and graph generative models with neural networks iteratively predict the next node connected to an existing node conditioned on the graph representing the current street layout, where street segments and intersections are respectively regarded as edges and nodes~\cite{2019_Chu}. Although these methods have demonstrated promising performances, obvious limitations can be observed about their adaptability to both natural and socioeconomic factors. In urban planning, the local terrain influences the expanding of streets considerably, and the distribution of population and the pattern of land use dominates travel demands that the street network aims to serve~\cite{1984_Sheffi}. The dependence of street layout design on natural and socioeconomic conditions has been incompletely considered in the current methods based on deep generative models, which is worth exploring.

To fill the above-mentioned gap, we propose a novel method for urban street layout design based on conditional adversarial learning, which is driven by both natural and socioeconomic data. In particular, our method first adopts an autoencoder~\cite{2011_Baldi} to encode the input rasterized data on elevation, population density and land use into a feature map; a conditional GAN (cGAN)~\cite{2014_Mirza, 2017_Isola} trained on the real-world dataset then takes as input the feature map and synthesizes street layout images in regard to the corresponding natural and socioeconomic conditions; and the following extraction module constructs and optimizes a street layout graph from each rasterized street layout for vectorization. Experiments and evaluations suggest that the proposed method can design street layouts resembling the street patterns in the corresponding urban regions with various details and demonstrating similar characteristics in terms of connectivity, street density, and traffic accessibility compared to their real-world counterparts. Meanwhile, the output street layouts are able to reflect the input conditions on elevation, population density and land use. Furthermore, we employ ArcGIS CityEngine~\footnote{https://www.esri.com/en-us/arcgis/products/arcgis-cityengine/overview} to create three-dimensional urban virtual environments from example outputs, which illustrates the effectiveness of our method.

The contributions of this work can be summarized as, (1) an end-to-end deep generative model for street layout image synthesis from fused natural and socioeconomic data; (2) an effective street layout extraction module for street layout graph generation from a synthesized street layout image; and (3) experiments and evaluations showing the proposed method's capability of designing various street layouts in regard to the input conditions while being similar to the real-world scenarios.

% To the best of our knowledge, CALL is the first method that incorporates population density, terrain and land use information in the street layout design process. We showcase the effectiveness of our approach by using CityEngine[5] to create a city model designed with CALL.

% ACM's consolidated article template, introduced in 2017, provides a
% consistent \LaTeX\ style for use across ACM publications, and
% incorporates accessibility and metadata-extraction functionality
% necessary for future Digital Library endeavors. Numerous ACM and
% SIG-specific \LaTeX\ templates have been examined, and their unique
% features incorporated into this single new template.

% If you are new to publishing with ACM, this document is a valuable
% guide to the process of preparing your work for publication. If you
% have published with ACM before, this document provides insight and
% instruction into more recent changes to the article template.

% The ``\verb|acmart|'' document class can be used to prepare articles
% for any ACM publication --- conference or journal, and for any stage
% of publication, from review to final ``camera-ready'' copy, to the
% author's own version, with {\itshape very} few changes to the source.
\section{Related Work}
% \paragraph{Traditional Methods} [1] presented a procedural approach to generate street layouts using L-System, which iteratively extended roads based on user-defined rules. This method was integrated into the commercial software CityEngine for practical use. Another approach proposed by [2] utilized traffic simulation and customization of random points to generate street layouts. [3] attempted to simulate the growth process of road networks by considering the interaction between population distribution and road network growth. Additionally, [4,5] focused on the impact of terrain on road network evolution, providing insights into the constraints imposed by terrain on street layout design.

\paragraph{Traditional Methods} \emph{Rule-based} methods have been widely used to solve the street layout design problem, given their strength in quick generation, especially since the pioneer work~\cite{2001_Parish} introduced~\emph{L-systems}~\cite{1986_Prusinkiewicz}, a forward procedural modeling technique originally for plant modeling, to urban modeling and had later developed into ArcGIS CityEngine for commercial use. Chen et al.~\cite{2008_Chen} applied tensor fields to the procedural modeling of street layouts. Vanegas et al.~\cite{2012b_Vanegas} presented an inverse procedural modeling system for generating plausible street layouts satisfying certain design indicators. Beneš et al.~\cite{2014_Benes} proposed a procedural modeling technique for growing local urban streets driven by simulating trade-based traffic with neighboring cities. Although rule-based methods enable compact and efficient street layout design with predefined grammars and insignificant storage, they usually require professional knowledge and expertise in relevant fields (e.g., urban planning, geography, and transport engineering). The parameters of grammars also need careful configuration by users for desirable outputs, which is uneasy to novices.

Some of earlier studies focus on \emph{example-based} methods for street layout design. Aliaga et al.~\cite{2008_Aliaga} proposed an urban layout synthesis algorithm based on the information provided by street intersections and aerial imagery. Nishida et al.~\cite{2016_Nishida} presented an interactive tool allowing the generation of street layouts with details and styles extracted from existing ones. Example-based methods ask for less user requirement compared to the rule-based regarding professional knowledge and expertise, but their outputs are excessively similar, sometimes even identical, to the source data, demonstrating the drawback on variety of design.

In the recent decade,~\emph{optimization-based} methods have gradually become a preferable solution to street layout design. Yang et al.~\cite{2013_Yang} integrated a global optimization module in their hierarchical splitting algorithm to ensure the fairness and regularity of the generated streets. Peng et al.~\cite{2016_Peng} proposed to create street layouts from high-level functional specifications with integer programming driven by topological constraints and scores. Feng et al.~\cite{2016_Feng} devised a crowd-aware layout design method based on stochastic optimization against crowd flow costs. Optimization-based methods can generate street layouts in an effective and uncomplicated way, but only if objective functions are appropriately formulated to represent user-given design goals and constraints. Otherwise, seeking optimal street layouts can be significantly difficult.

\begin{figure*}[t]
    \centering  
    \includegraphics[width=1\textwidth]{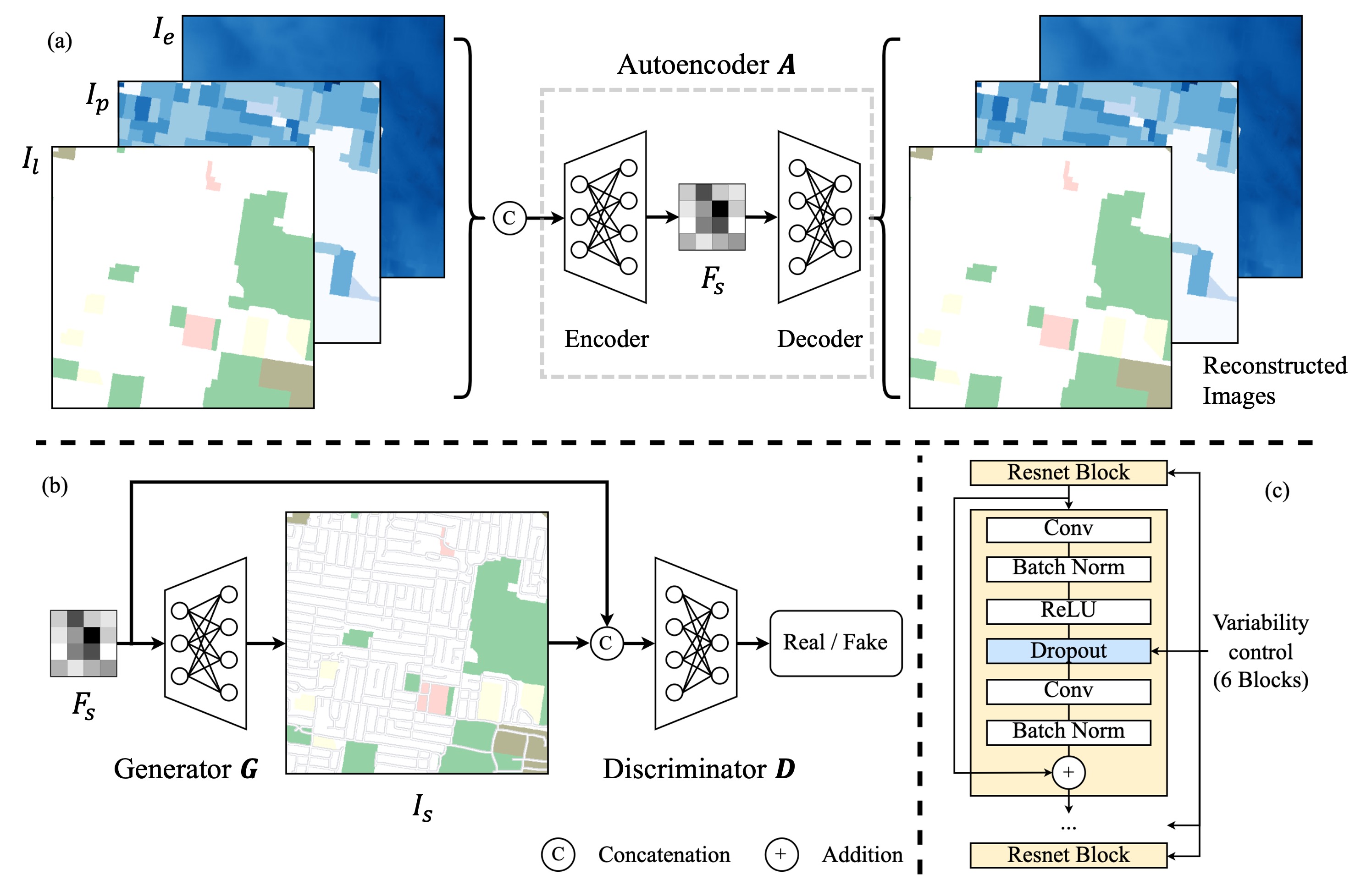}    
    \caption{Architecture of the street layout image synthesis module. (a) The autoencoder~$\mathcal{A}$ encodes the input images $I_{e}$, $I_{p}$, and $I_{l}$ into a feature map $F_{s}$. (b) The generator~$\mathcal{G}$ maps $F_{s}$ to a street layout image $I_{s}$, and the discriminator~$\mathcal{D}$ distinguishes the images corresponding the real-world street layouts from the synthesized. (c) The structure of~$\mathcal{G}$ comprises includes dropout layers for controlling the variability of the synthesized images.}
    % \Description{Enjoying the baseball game from the third-base seats. Ichiro Suzuki preparing to bat.}
    \label{fig:arch}
\end{figure*}

\paragraph{Deep Generative Models.} Recent methods based on deep generative models solve the problem following diverse strategies.~\emph{Graph-based} methods characterize a street layout as a planar graph comprising nodes and edges, and thus formulate the problem as graph generation. Chu et al.~\cite{2019_Chu} proposed Neural Turtle Graphics (NTG), based on an encoder-decoder architecture where the encoder recurrent neural network (RNN) encodes incoming paths into a node and the decoder RNN creates outgoing nodes and connecting edges. Despite the success in learning local street patterns, NTG can hardly process large-scale street layouts. Owaki et al.~\cite{2022_Owaki} improved the encoder and decoder neural networks in NTG to predict the attributes of both streets and blocks. These methods concentrate on the topology of a street layout as graph, whereas the proposed method aims to design street layouts adapted to natural and socioeconomic conditions.

The other \emph{image-based} methods formulate street layout design as image synthesis. Hartmann et al.~\cite{2017_Hartmann} firstly proposed StreetGAN, a GAN-based method synthesizing textures to represent street layouts from a noise image, which has a limited generative capability and is prone to model collapse~\cite{2016_Mao,2017_Arjovsky}. Fang et al.~\cite{2020_Fang} devised an image complement method predicting street layouts having a limited size in a user-specified region. More recently, Kelvin et al.~\cite{2020_Kelvin} introduced a cGAN-based tool for generating street map tiles from a user-defined sketch, which however lacks details on the output quality. Fang et al.~\cite{2021_Fang,2022_Fang} devised cGAN-based methods for street layout design given planning guidance defined by users and contextual information closely related to topography. Although our method is similarly cGAN-based, it employs data fusion to simultaneously integrate the conditions on elevation, population density, and land use in the design loop, which echos the rule in urban planning that street layout design requires the contemplation of natural and socioeconomic factors.

% CALL is able to produce street layouts that accurately reflect the underlying population distribution in each location, while also considering terrain and land use constraints. 

% CALL's ability to generate street layouts that are both interpretable and customizable make it well-suited for a variety of applications.

\section{Overview}

The proposed method comprises a \emph{synthesis} module and an \emph{extraction} module. The synthesis module adopts an autoencoder to produce a feature map $F_{s}$ by fusing three single-band images $I_{e}$, $I_{p}$, and $I_{l}$ that project the information on elevation, population density, and land use, respectively, of a region of interest $\mathcal{R}$. Subsequently, a cGAN trained on a real-world dataset takes as input $F_{s}$ and synthesizes a single-band image $I_{s}$, which represents a street layout in raster data (see Section~\ref{sec:sl_syn}).

The extraction module initially processes the synthesized image $I_{s}$ into an enhanced image $\hat{I}_{s}$ that filters the unwanted artifacts. Afterwards, a planar graph ${G}_{s}$ is constructed from $\hat{I}_{s}$ by comparing each pixel and its neighbors, followed by optimizing it into the final output graph $\hat{G}_{s}$, which represents a street layout in vector data (see Section~\ref{sec:sl_ext}).

Once $\hat{G}_{s}$ meets the user's expectation, it can be easily converted into a standard geospatial data format (e.g., SHAPEFILE\footnote{https://doc.arcgis.com/en/arcgis-online/reference/shapefiles.htm}) and thus serves three-dimensional urban virtual scene creation using a mainstream urban modeling platform (e.g., ArcGIS CityEngine). Otherwise, our method can regenerate street layouts with updated details but still corresponding to the same input data for user choice.

\section{Street Layout Image Synthesis}
\label{sec:sl_syn}

The synthesis module aims to synthesize street layouts from natural and socioeconomic conditions, where all these data are represented as images. Therefore, our synthesis task can be formulated as predicting an output $y$ from an input $x$.

Inspired by a two-player zero-sum minimax game, GAN~\cite{2020_Goodfellow} is a deep generative models comprising a generator $\mathcal{G}$ and a discriminator $\mathcal{D}$. Focusing on the mapping from a random noise $z$ to a \emph{fake} image $\hat{y}$ (i.e., $\hat{y}=\mathcal{G}(z)$), $\mathcal{D}$ is trained to differentiate a \emph{real} image $y$ from $\hat{y}$, whereas $\mathcal{G}$ aims to synthesize $\hat{y}$ indistinguishable to $\mathcal{D}$. To incorporate other input information, cGAN~\cite{2014_Mirza} enables a slightly different process: its $\mathcal{G}$ synthesizes $\hat{y}$ from a user-given condition $x$ in addition to $z$ (i.e., $\hat{y}=\mathcal{G}(x,z)$). In this way, the input-output relationship can be explored to support controlled generation.

Our street layout synthesis module~$\mathcal{S}$ is cGAN-based, as following the recent practices on terrain synthesis~\cite{2017_Isola}, street map tiles generation~\cite{2020_Kelvin}, and topography-aware street network modeling~\cite{2022_Fang}. Unlike the original cGAN, our $\mathcal{S}$ takes as input only a condition set $x=\{I_{e}, I_{p}, I_{l}\}$, but without any random noise $z$. This design follows our observation that random noise usually leads to the degeneration of the output quality, because the model may simply learn how to remove the noise rather than extract the underlying relationship serving our task, which is consistent with the findings of previous studies~\cite{2015_Mathieu,2017_Isola}. In the dataset for training $\mathcal{S}$, each condition set $x=\{I_{e}, I_{p}, I_{l}\}$ corresponds to an image $y$ representing a real-world street network, which enables a one-to-one mapping. The limited number of samples available for each condition set, however, result in that $\mathcal{S}$ may struggle to learn both stable structural features and textural differences from real images.

As an image contains limited variation but considerable redundant information and even noise, we employ an autoencoder $\mathcal{A}$ to process the input images $I_{e}$, $I_{p}$, and $I_{l}$ projecting the information on elevation, population density, and land use, given its outstanding capability of compressing and downscaling data~\cite{2006_Hinton,2011_Baldi}. Specifically, $\mathcal{A}$ is composed of an encoder, which compresses $I_{e}$, $I_{p}$, and $I_{l}$ into a feature map $F_{s}$, and a decoder, which reconstructs the images from the feature map. This process of data compression can be formulated as,
\begin{equation}
F_{s} = \mathcal{A}(I_{e}, I_{p}, I_{l}).
\end{equation}

Conventional GANs usually construct the discriminator $\mathcal{D}$ as a classifier trained with the Sigmoid cross-entropy loss, which sometimes causes vanishing gradients and leads to instability and unsatisfactory quality in synthesized images~\cite{2017_Arjovsky}. To address this issue, we adopts a least-squares loss~\cite{2016_Mao} to train the generator $\mathcal{G}$, whose objective function is formulated as,
\begin{equation}
min\mathcal{L}_\mathcal{G}=\mathbb{E}_{F_s}[(\mathcal{D}(\mathcal{G}(F_s),F_s){-1)}^2],
\end{equation}
and the objective function of $D$ is formulated as,
\begin{equation}
min\mathcal{L}_D=\frac{1}{2} \mathbb{E}_{x,F_s}[(\mathcal{D}(x,F_s){-1)}^2]+\frac{1}{2}\mathbb{E}_{F_s}[(\mathcal{D}(\mathcal{G}(F_s),F_s))^2].
\end{equation}

Figure~\ref{fig:arch} shows the architecture of our street layout image synthesis module, which comprises an autoencoder~$\mathcal{A}$, a generator~$\mathcal{G}$, and a discriminator~$\mathcal{D}$. We adopt convolution-BatchNorm-ReLu~\cite{2015_Radford} network blocks and residual blocks~\cite{2015_He} with a dropout layer~\cite{2014_srivastava}, as illustrated in Figure~\ref{fig:arch}~(c).

\subsection{Autoencoder}
Our autoencoder~$\mathcal{A}$ compress the concatenated input images $I_{e}$, $I_{p}$, and $I_{l}$ of size $512\times 512\times 3$ into a feature map $F_{s}$ of size $64\times 64$, which captures salient image features. Specifically, the encoder includes an input layer, downsampling blocks, residual blocks, and an output layer, where the downsampling blocks progressively reduce the dimensionality and size of the input and the residual blocks enhance the network's ability to extract useful features; and the decoder is structurally symmetrical to the encoder, comprising an input layer, residual blocks, upsampling blocks, and an output layer. The encoder's output layer produces $F_{s}$. Our autoencoder~$\mathcal{A}$ enables efficient compression of the input images while preserving their critical information for street layout image synthesis.

\subsection{Generator}
Our generator~$\mathcal{G}$ takes as input the feature maps $F_{s}$ without any random noise, and includes an input layer, downsampling blocks, residual blocks, upsampling blocks, and an output layer. As dropout layers can be used to prevent overfitting and improve the network's generalization~\cite{2014_srivastava}, we introduce dropout layers to six residual blocks and two upsampling blocks, which enables $\mathcal{G}$ to structurally or texturally control the synthesized image $I_{s}$. In particular, a street layout's structure and connectivity can be controlled by randomly masking a portion of high-level semantic features in the residual blocks and of low-level semantic features in the upsampling blocks, respectively. By avoiding the use of random noise and introducing dropout layers, our method can synthesize high-quality street layout images with various structures and connectivities.

% During training, we set the dropout layer parameters to 0.5. In the experimental section, we showcase the diverse generative capabilities of our model in detail. By avoiding the use of random noise and incorporating dropout layers, our model produces high-quality street layout images with varying road structures and connectivity.

\subsection{Discriminator}
Reasonable streets should be geometrically continuous lines or curves, which are represented as high-frequency features in a street layout image. Conventional GAN discriminators evaluate the average quality of images at a global level, which may cause image blurring, or even training failure. In contrast, our discriminator~$\mathcal{D}$ follows the design of PatchGAN~\cite{2017_Isola} that focuses on patch-scale structures for fine-grained evaluation of image quality, which ensures the quality of synthesized street layout images.

% Specifically, our discriminator~$\mathcal{D}$ performs binary classification for each $N \times N$ patch of the input images. Our experimental results demonstrate the effectiveness of this approach, which enhances the quality of generated street layout images.

\begin{figure*}
    \centering  
    \includegraphics[width=0.9\textwidth]{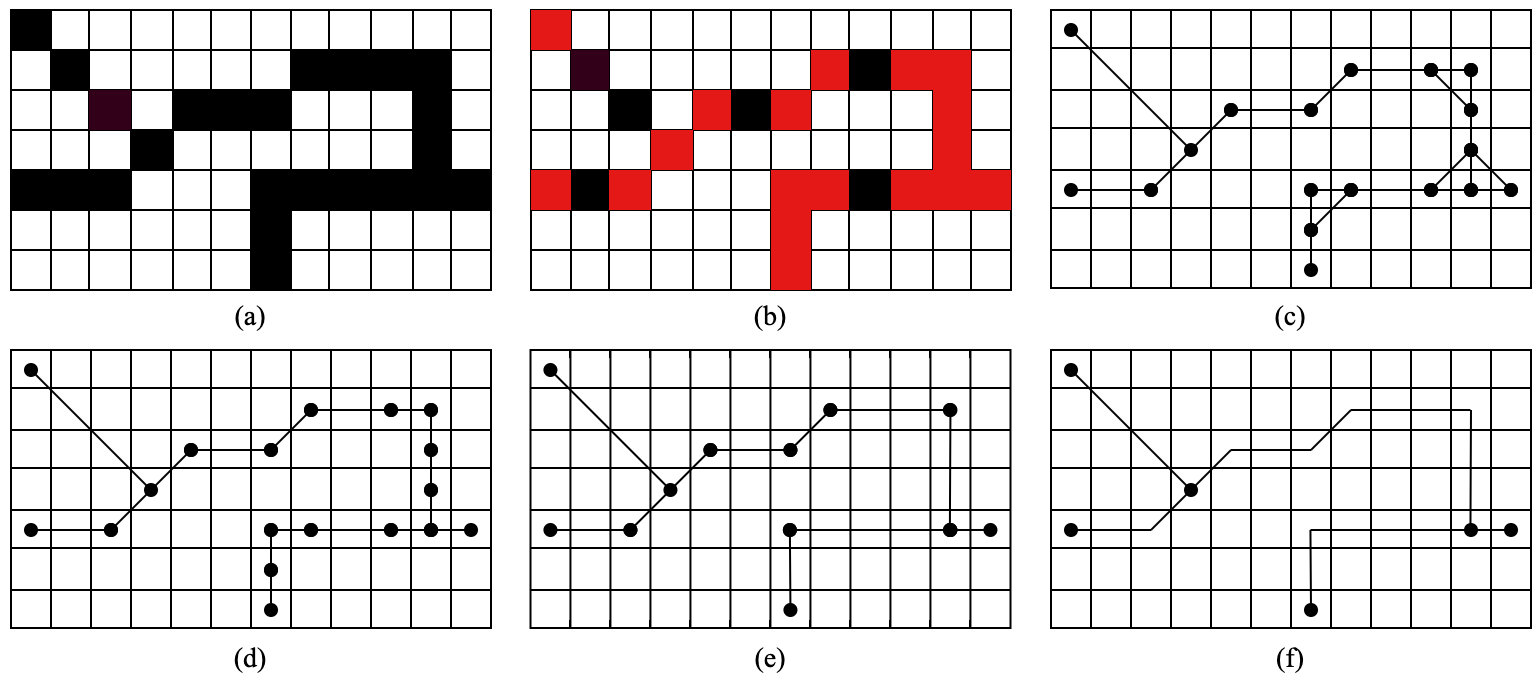}    
    \caption{Process of the street layout graph extraction. (a) The initial street layout skeleton, where streets are represented in black and non-street areas in white. (b) The street layout skeleton after vertex selection, where invalid vertices are in red. (c) The constructed street layout graph $G_{s}$, where isosceles right triangles exist at street corners. (d) The street layout graph after the removal of isosceles right triangles. (e) The street layout graph after the removal of redundant vertices. (f) The optimized street layout graph $\hat{G}_{s}$.}
    % \Description{Enjoying the baseball game from the third-base seats. Ichiro Suzuki preparing to bat.}
    \label{fig:ext}
\end{figure*}

% \subsection{Training and Synthesis}
% Prior to training the synthesis module~$\mathcal{S}$, we pre-train the autoencoder~$\mathcal{A}$ by optimizing the L2 loss between the original input images and the reconstructed with the Adam solver~\cite{2014_Kingma}. The initial learning rate is set 0.0002 and momentum parameters $\beta_1 = 0.5$ and $\beta_2 = 0.999$. We select the best performing Autoencoder based on its performance on a separate test dataset. We then use the encoder and freeze all its parameters for training cGAN. 

% In the cGAN training phase, we alternate between one gradient descent step on the discriminator and one step on the generator. During training, we set the parameters of the dropout layer to 0.5. We use the Adam solver with a learning rate of 0.0002 for both the generator and discriminator, momentum parameters $\beta_1 = 0.5$ and $\beta_2 = 0.999$. We update the learning rate every 300 epochs with a weight of 0.1.

% {\bfseries Inference.} During the inference phase, we utilize our trained encoder and generator models to generate street layout images. Unlike traditional approaches, we incorporate dropout layers into our testing procedure to enhance the generated image quality. Specifically, we experiment with varying the parameters of the dropout layer between 0.5 and 0.95 to investigate their impact on the generated images. By randomly dropping out certain features during training, the network is forced to rely on other features, resulting in more robust and diverse image generation.

\section{Street Layout Graph Extraction}
\label{sec:sl_ext}

The output of the street layout image synthesis module is a grayscale image~$I_{s}$ by the generator~$\mathcal{G}$ from the feature map ~$F_{s}$, which needs converting to a planar graph for downstream tasks, such as statistical analysis, urban modeling, and data visualization. Our street layout graph extraction module includes three steps: image processing, graph construction, and graph optimization.

\subsection{Image Processing}
We first convert the street layout image~$I_{s}$ to a binary image using a threshold value of $127$, where streets are represented by pixels in white with value of $255$ and other pixels in black with value of $0$ correspond to non-street areas. The following morphological expansion operation aims to remove artifacts, including tiny holes, disconnected segments, and parallel streets, and produces a cleaned binary image for street layout skeleton extraction~\cite{1984_Zhang}. Iterative operations of morphological expansion and skeleton extraction can enhance the quality of the street layout skeleton in the processed image~$\hat{I}_{s}$ without compromising the structural integrity. Please note that the colors adopted in Figure~\ref{fig:ext} are for visualization only, where streets are in black and non-street areas in white.

\subsection{Graph Construction}
From the processed image~$\hat{I}_{s}$ representing a street layout skeleton (see Figure~\ref{fig:ext}~(a)), we construct a planar graph $G_{s} = \{V, E\}$ with a vertex set $V = \{v_i\}$ and an edge set $E = \{e_{ij}\}$. Each vertex $v_i$ encodes a spatial location $[x_i, y_i]^\top$, and each edge $e_{ij} = \overline{v_iv_j}$ denotes a street segment. We define a pixel with value of $255$ as valid, and check the $8$ neighboring pixels for each valid pixel. Each valid pixel $p_i$ will be regarded as a vertex $v_i$ added to $V$ if its neighboring pixels meets one of the following criteria:
\begin{itemize}
\item the existence of $1$ valid pixel;
\item the existence of more than $2$ valid pixels; and
\item the existence of $2$ valid pixels, to which $p_i$ cannot be connected in a straight line.
\end{itemize}

We traverse~$\hat{I}_{s}$ to find all valid vertices that satisfying any of the above criteria (see Figure~\ref{fig:ext}~(b)). For each vertex $v_i$, we locate its adjacent vertices $v_j$ and add an edge $e_{ij}=\overline{v_iv_j}$ to $E$, which produces an initial street layout graph $G_{s}$.

\subsection{Graph Optimization}
The constructed street layout graph $G_{s}$ may exhibit undesirable isosceles right triangles at street corners, as shown in Figure~\ref{fig:ext}~(c). We find each of such triangles by traversing the vertices and then remove its longest edge (see Figure~\ref{fig:ext}~(d)). Besides, the intermediate graph usually contains a substantial number of redundant vertices hardly contributing to the street layout. For simplification, we eliminate such vertices (see Figure~\ref{fig:ext}~(d)) and merge corresponding edges to obtain the final street layout graph $\hat{G}_{s}$, as shown in Figure~\ref{fig:ext}~(f).

\section{Experiments} 
We implemented the proposed method using Python and PyTorch on a workstation with a NVIDIA RTX A5000 graphics card (24 GB memory), and created three-dimensional urban virtual environments for showcase using ArcGIS CityEngine on a laptop with a NVIDIA GeForce RTX 3080Ti graphics card (12 GB memory). Evaluations were all conducted on the street layout graphs. For visualization, we assigned colors to the street layouts according to the land use types.

\subsection{Implementation Details}
To train the synthesis module~$\mathcal{S}$, we pre-trained the autoencoder~$\mathcal{A}$ by optimizing the $L2$ loss between the original input images and the reconstructed ones using the Adam solver~\cite{2014_Kingma} with momentum parameters $\beta_1 = 0.5$ and $\beta_2 = 0.999$. The learning rate was set to $0.0002$ and decayed to one-tenth for every $300$ steps. We selected the best autoencoder by performance on an independent test set. Then, we used the encoder while freezing all of its parameters to train the cGAN. During the training of the cGAN, we alternated between one gradient descent step on the discriminator~$\mathcal{D}$ and one step on the generator~$\mathcal{G}$. The dropout rate in~$\mathcal{D}$ was $0.5$, where $\mathcal{D}$ performed binary classification for each $64 \times 64$ patch of the input images. 

During the testing, we adopted the encoder of $\mathcal{A}$ and $\mathcal{G}$ to synthesize street layout images. Unlike traditional methods, we incorporated dropout layers in testing to enhance the image quality using the dropout rate ranging from $0.5$ to $0.99$. We found that the dropout rate of $0.95$ corresponded to street layout images with stable structural integrity and textural variability.

\subsection{Dataset Preparation}
We produced a real-world street layout dataset based on the map data from OpenStreetMap (OSM)~\footnote{https://www.openstreetmap.org}. Specifically, each street in the OSM map data is tagged with a specific~\emph{highway} type representing its category. We only selected streets tagged~\emph{highway}-categories:~\emph{motorway},~\emph{primary},~\emph{secondary},~\emph{tertiary},~\emph{residential},~\emph{living street}, or ~\emph{pedestrian}, from three major Australian cities (i.e., Sydney, Melbourne, and Brisbane). Due to incomplete data tagging in those underpopulated regions, we only adopted the map data for the populated regions to ensure adequate representation.

The map data (i.e., vectorized street layouts) in SHAPEFILE was rasterized using the OSGEO library~\footnote{https://gdal.org/api/python/osgeo.html}, involving the projection of the street layouts into the Web Mercator coordinate system in meters. All street layouts images were cropped to $512\times 512$ patches with a $5$m spatial resolution and street width of $3$ pixels. In particular, streets were represented by pixels in white with value of $255$, and non-street areas by pixels in black with value of $0$. In addition, we rasterized the 2021 Census data~\footnote{https://www.abs.gov.au/census} into $512\times 512$ images that projected the information on elevation, population density and land use corresponding to the collected street layouts. Our dataset consists of $2586$ pairs of street layout images and natural / socioeconomic images, of which 70\% were for training and 30\% for testing.

\begin{figure}[t]
  \centering
  \includegraphics[width=\linewidth]{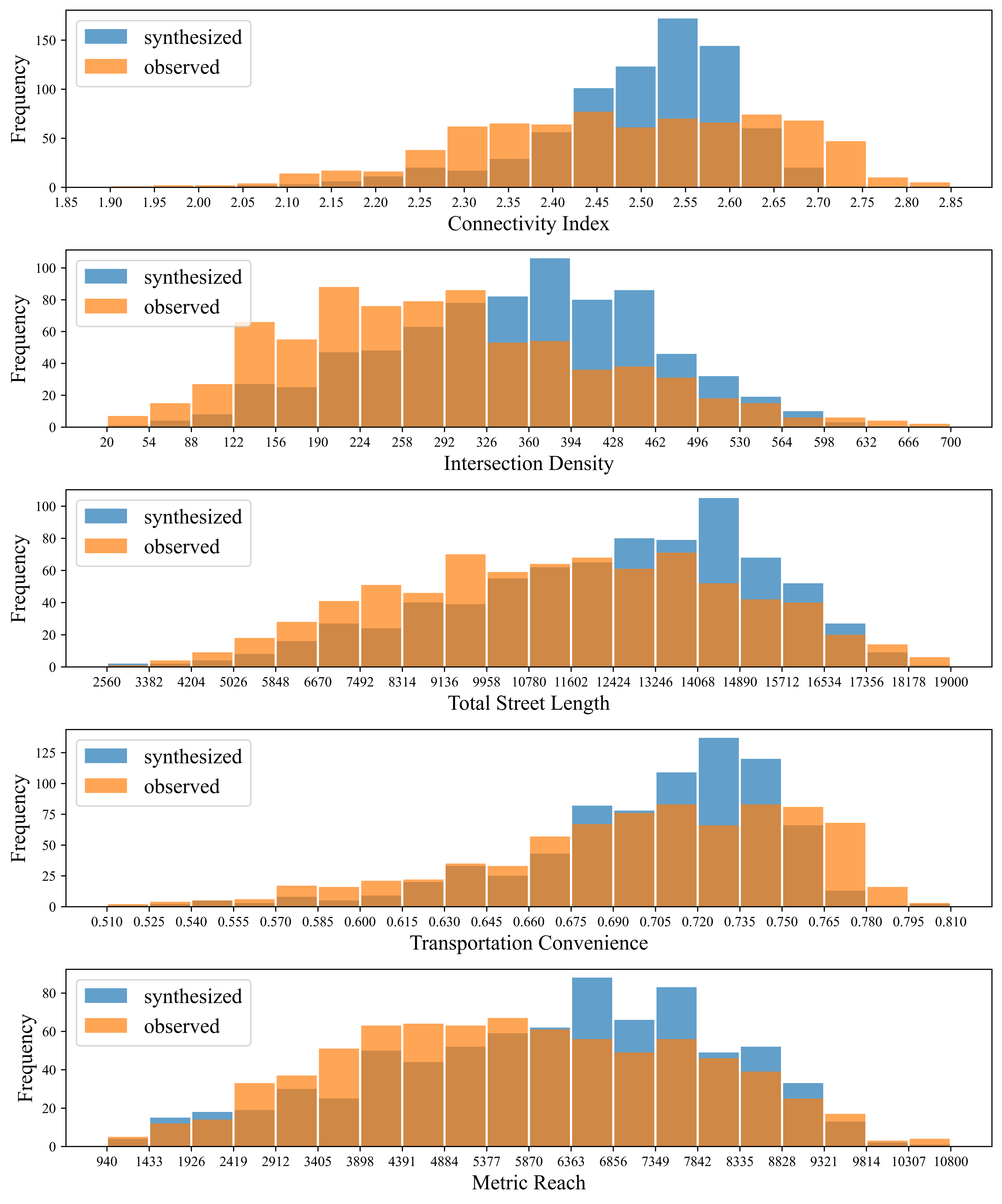}
  \caption{Comparison between the street layouts output from the proposed method (blue) and the real-world ones (orange) in statistical evaluation.}
  % \Description{A woman and a girl in white dresses sit in an open car.}
  \label{fig:sta}
\end{figure}

\subsection{Statistical Evaluation}
In the field of urban planning, numerous studies have focused on street layout qualitative evaluation regarding the capability to support various urban functions (e.g, pedestrian and vehicular movements). As an urban street layout can be represented as a planar graph, the topological and geometric attributes may resemble its quality and pattern. For example, the central area of a city with a radial layout is supposed to exhibit high connectivity and provide easy access, whereas residential communities are generally located around dead-end streets or T-junctions to reduce traffic for privacy and security. 

\begin{figure}[t]
  \centering
  \includegraphics[width=\linewidth]{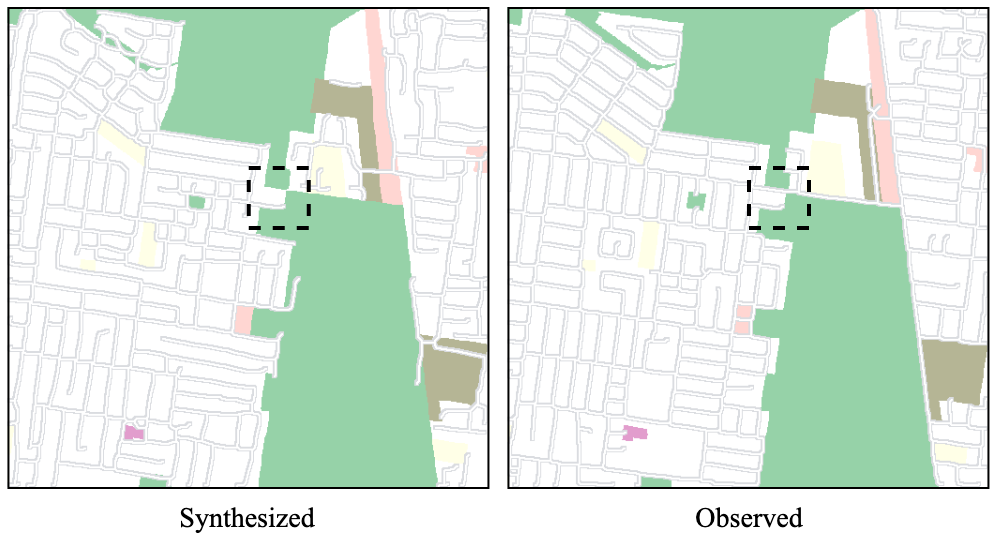}
  \caption{Comparison between an example street layout output from the proposed method (left) and its corresponding real-world counterpart (right).}
  % \Description{A woman and a girl in white dresses sit in an open car.}
  \label{fig:com}
\end{figure}

\begin{figure*}[t]
    \centering  
    \includegraphics[width=0.95\textwidth]{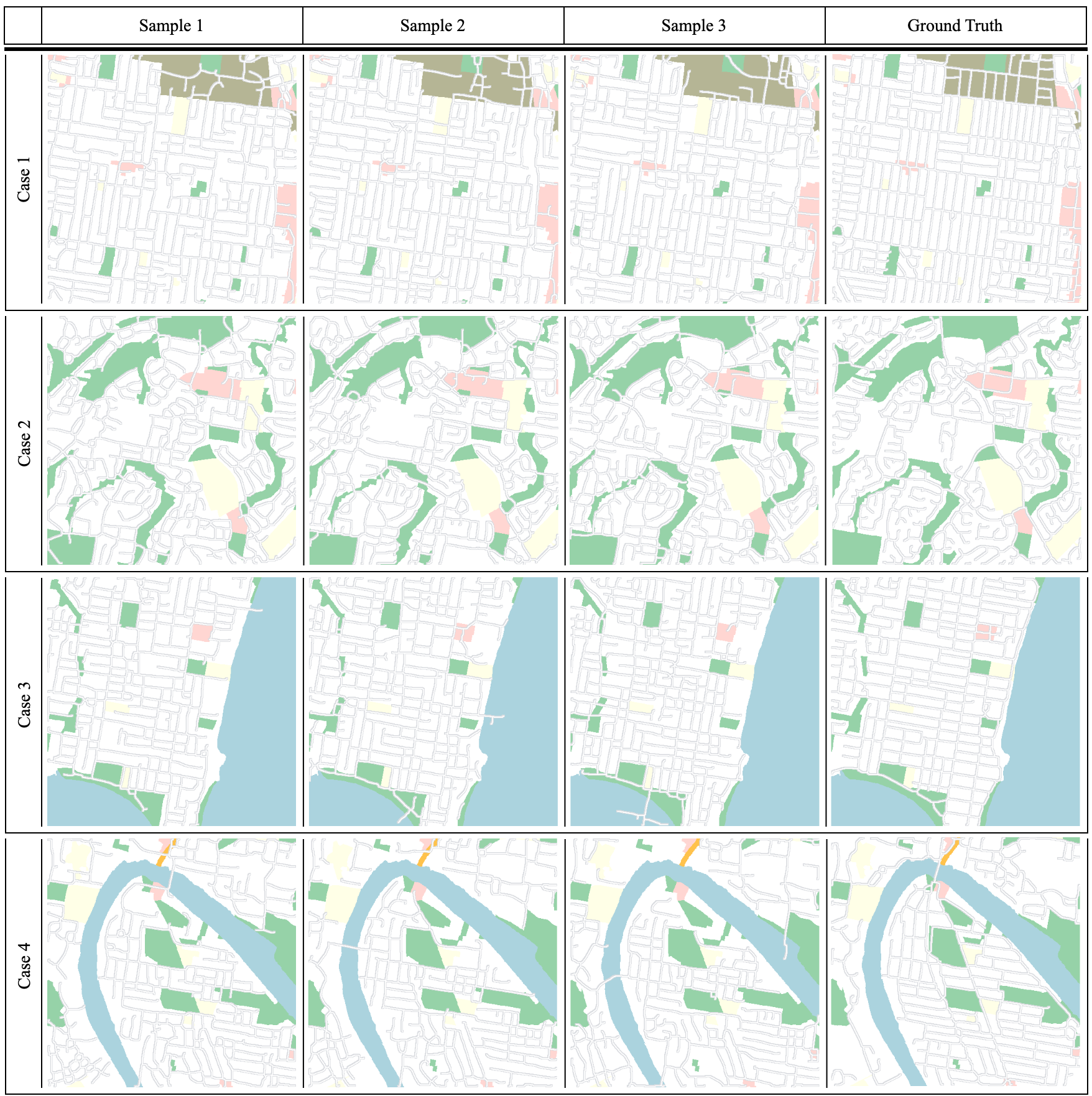}    
    \caption{Visualization of the street layouts outputs output from the proposed method and the real-world ones in $4$ cases. Land use types are identified (brown: industrial; green: parkland; red: commercial; yellow: education; white: residential; orange: transport; blue: water).}
    % \Description{Enjoying the baseball game from the third-base seats. Ichiro Suzuki preparing to bat.}
    \label{fig:vis}
\end{figure*}

Therefore, we adopted the following the statistical metrics~\cite{2014_Sawsan} to assess the topological and geometric attributes of a street layout regarding connectivity and pattern encoding. 

\paragraph{Connectivity Index ($t_{ci}$)} We defined the connectivity index as the average vertex degree of the corresponding graph to measure the quality of a street layout in connecting destinations, which is formulated as,
\begin{equation}
t_{ci}(G)=\frac{\sum_{i} degree(v_i)}{number(v_i)}.
\end{equation}
In particular, we focused on vertices with degrees other than $2$. It is noteworthy that urban planning guidelines usually employ a minimum connectivity index of $1.4$~\cite{1984_Sheffi}, which underscores the importance of ensuring adequate connectivity in a street layout to support efficient transportation and facilitate access to key destinations.

\paragraph{Intersection Density ($t_{id}$)} The metric of intersection density serves as a quantitative measure of the number of intersections present within a given patch layout, defined as the total number of vertices with degrees greater than $2$. This metric provides valuable insights into the connectivity and accessibility of a street layout and can inform decision-makers of either optimizing the efficiency and functionality of transportation infrastructure.

\paragraph{Total Street Length ($g_{sl}$)} The effectiveness of transportation infrastructure is intrinsically linked to the total length of streets. Areas with high population density, for instance, often exhibit densely distributed streets, while rural areas are typically characterized by a sparse network.

\paragraph{Transportation Convenience ($g_{tc}$)} Transportation convenience measures the easiness of traveling from source $s$ to destination $d$. Suppose $d_E(s, d)$ refers to the Euclidean distance between $s \rightarrow d$ and $d_D(s, d)$ to the shortest graph distance, the transportation convenience between $s$ and $d$ is formulated as,
\begin{equation}
g_{tc}(s,d)= \frac{d_E(s,d)}{d_D(s,d)}.
\end{equation}
For a street layout, we randomly sampled 100 sets of source and destination locations and calculated the average transportation convenience. To avoid local estimation, source and destination locations were required to have a minimal Euclidean distance ($d_E(s, d) > 0.25km$ in our implementation).

\paragraph{Metric Reach ($g_{mr}$)} As a well-established concept in urban planning literature, metric reach is defined as the total length of streets that can be accessed by traveling a given distance ($0.5km$ in our implementation) in all possible directions, without retracing any paths, which provides a comprehensive assessment of the accessibility and connectivity of a street layout.

Figure~\ref{fig:sta} compares statistical evaluation results between the synthesized street layouts by our method and the real-world street layouts. In all metrics, the synthesized street layouts demonstrate that the distribution of the synthesized street layout is consistent with the real-world ones. Specifically, our model prefers to synthesize streets with high connectivity and numerous intersections, which can be derived from the connectivity index and intersection density, where larger connectivity index represents more streets connected to an intersection. This indicates that the synthesized street layout is able to support a heavy travel demand and maintain efficient transportation. From total street length and metric reach it can be concluded that the synthesized street layout has denser streets and better accessibility compared to the real-world. The synthesized street layouts do not outperform the real-world ones in terms of high transportation convenience. Further analysis reveals that some of the synthesized street layouts lack a bridge-like structure, as shown in Figure~\ref{fig:com}. Specifically, the synthesized street layout is divided into two separate pieces on the left and right. Consequently, we observe a reduction in both connectivity index and transportation convenience. In summary, our model demonstrates its ability to learn the geometric and topological properties of the real-world street layout.

\subsection{Visual Evaluation}
\begin{figure}[t]
  \centering
  \includegraphics[width=\linewidth]{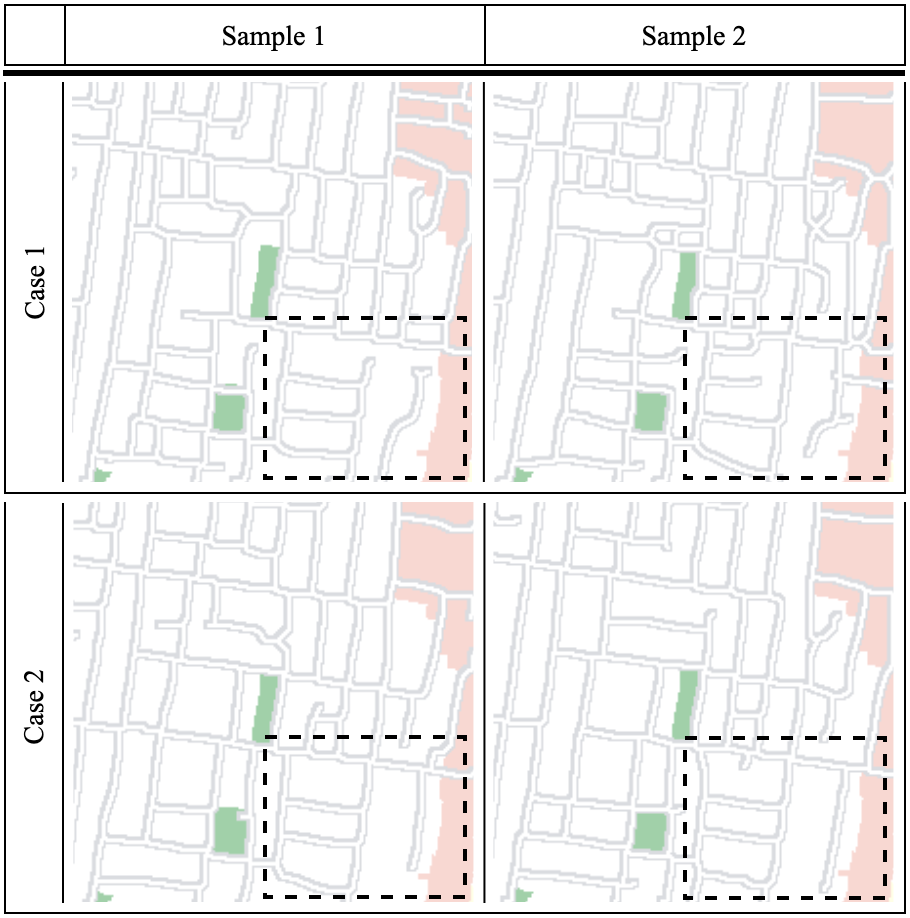}
  \caption{Visualization of variability control by the proposed method. Land use types are identified (brown: industrial; green: parkland; red: commercial; yellow: education; white: residential. The black dotted box is the interest area).}
  % \Description{A woman and a girl in white dresses sit in an open car.}
  \label{fig:var}
\end{figure}

We also evaluated the generative capability of our method from a visual perspective. In each of the four cases, we synthesized $1000$ street layouts and selected three of them for presentation (see Figure~\ref{fig:vis}). We controlled the variability of the synthesized street layouts by setting the dropout rate. In particular, we adopted the dropout rate of 0.95 in the $6$ network blocks, which could maintain the image quality and ensure the variability (see Figure~\ref{fig:var}). 

\paragraph{Case 1} As a grid-like street layout, it is characterized by containing rectangular patterns of varying sizes, like a chessboard. It can be noticed that the synthesized street layout is similar in structure but has a varied texture compared to its real-world counterpart. In particular, residential areas have more rectangular streets than those of other land use types.

\paragraph{Case 2} As a curved street layout, it contains a lot of curved streets and dead ends. It can be noted that the synthesized street layout reproduces the curved streets of high quality. In addition, it captures the different land use types accurately. The large parkland area indicates a strong correlation between the synthesized street layout and the input population density. Specifically, the streets are significantly sparse on a parkland area, as there are almost no residents.

\paragraph{Case 3} As a street layout surrounded by water, no streets are placed over the water, whereas the synthesized street layout has a grid-like pattern alike its real-world counterpart.

\paragraph{Case 4} As a street layout including river, it can be observed that the synthesized street layout is divided into two parts with bridges generated to maintain connectivity. In addition, the streets are deliberately bypass parkland areas and the river.

In addition, we changed the position, number and parameters of the dropout layer in the network blocks to control the variability of the synthesized street layouts, as shown in Figure~\ref{fig:var}. For Case 1, we added $6$ dropout layers to high-level features, randomly hiding $95\%$ of the feature information, which led to a structural change in the street layout. For Case 2, we added $1$ dropout layer to the low-level features, randomly hiding $95\%$ of the feature information, which led to a slight change in the texture of the street layout. 

The street layouts output from our method can reflect the conditions on elevation, population density and land use. We used ArcGIS CityEngine to create three-dimensional urban virtual environments from the the synthesized street layouts, illustrating the effectiveness of our method (see Figure~\ref{fig:teaser}).

\section{Conclusion}
In this paper, we proposed an end-to-end deep generative model for street layout image synthesis from fused natural and socioeconomic data, and an effective street layout extraction module generating corresponding graphs. Experiments and evaluations suggested that our method could generate various street layouts representing the input conditions, while being similar to the real-world cases. Furthermore, we utilized the output street layouts for three-dimensional urban virtual environment creation, which showed the method's promise in urban modeling. In future work, we are to explore the design of street layouts with increased properties (e.g., levels and modes).

\bibliographystyle{ACM-Reference-Format}
\bibliography{sample-base}

% %%
% %% If your work has an appendix, this is the place to put it.
% \appendix









\end{document}